\documentclass{article}
\usepackage{amssymb}

\input{tcilatex}

\begin{document}

\title{PROPERTIES OF EXCITED STATES\\
IN THE $^{160}$Dy NUCLEUS}
\author{J. Adam$^{1,2}$, V. P. Garistov$^{3}$,M. Honusek$^{2}$, \\
J. Dobes$^{2}$, J. Zvolski$^{2}$, J. Mrazek$^{2}$, A.A. Solnyshkin$^{1}$ \\
$^{1}${\small Joint Institute for Nuclear Research, Dubna, Russia.}\\
$^{2}${\small \ Institute of Nuclear Physics, Czech Academy of Sciences, Rez.%
} \\
$^{3}$ {\small Institute for Nuclear Reseach and Nuclear Energy,BAS, Sofia,
Bulgaria}}
\maketitle

\begin{abstract}
Positive-parity levels and 16 rotational bands are theoretically analyzed on
the basis of phenomenological models of the atomic nucleus with the use of
new experimental data on excited states in the $^{160}$Dy nucleus recently
gained in the investigation of the decay $^{160}$Er $\longrightarrow $ $%
^{160m,g}$Ho $\rightarrow $ $^{160}$Dy.
\end{abstract}

{\LARGE INTRODUCTION \ }

\bigskip The $^{160}$Dy nucleus is classified with deformed nuclei $(\beta
=0.23)$ and has quite a complicated scheme of excited states. By now it has
been well studied experimentally in nuclear reactions, Coulomb excitation,
and $\beta $ decays of $^{160}Tb$ and $^{160m,g}Ho$ [1]. Our recent\
investigation of the decay $^{160}Er$ $\rightarrow $ $^{160m,g}Ho$ $%
\rightarrow $ $^{160}Dy$ [2]\bigskip\ has made it possible to expand
considerably the scheme of excited $^{160}Dy$ states and to correlate the
reaction and $\beta $ decay data. Over a hundred new levels are added to the
previously known excited states in the $^{160m,g}Ho$ $\rightarrow $ $%
^{160}Dy $ decay scheme. The complete list of these levels is given in Table
1 together with their quantum characteristics. In this paper positive-parity
level energies calculated by us are compared with the experimental data and
the experimentally known positive-parity and negative-parity bands
(sometimes with new levels added by us) in the $^{160}Dy$ nucleus are
theoretically analyzed on the basis of existing nuclear models.

\section{ POSITIVE PARITY LEVELS}

Properties of deformed nuclei can be described within the symmetrical
rotator model [3] and the interacting boson model IBM-1 [4] in the $SU(3)$
limit. Some differences in the descriptions are due to the symmetry of the
Hamiltonian and the finite number of bosons which only have the angular
momentum $L=0$ and $L=2.$ In the IBM-1 we used a simple Hamiltonian 
\begin{equation}
H=-kQQ-k^{\prime }LL+k^{\prime \prime }PP.
\end{equation}

In the $SU(3)$ limit $k^{\prime \prime }=0,$ the energies of the $\beta $
and $\gamma $ bands are degenerate and become identical, the energies of the
ground-state band are designated as $(\lambda ,0)$, those of the $\beta $
and $\gamma $ bands as $(\lambda -4,2),$ and these energies are calculated
as a function of the spin $I$ by the formula%
\begin{equation}
E(I)=(0.75k-k^{\prime })I(I+1)-kC(\lambda ,\mu ),
\end{equation}

where $C(\lambda ,\mu )$ are the eigenvalues of the Casimir operator

\begin{equation}
C(\lambda ,\mu )=\lambda ^{2}+\mu ^{2}+\lambda \mu +3(\lambda +\mu ),
\end{equation}

here $\lambda $ is the number of valence nucleons (for $^{160}Dy$ we have $%
\lambda $ $=28$). Using equations (2) and (3) and the experimental energies
of the first $(86.8keV)$ and second $(966.2keV)$ $2^{+}$ levels in the $%
^{160}Dy$ nucleus, we calculate the coefficients $k$ and $k^{\prime }$ by
the formulae 
\begin{equation}
k=\frac{E(2_{2}^{+})-E(2_{1}^{+})}{6(\lambda -1)}
\end{equation}

\begin{equation}
k^{\prime }=0,75k-\frac{E(2_{1}^{+})}{6}
\end{equation}%
and get $k=5.43keV,$ $k^{\prime }=10.39keV$. Since the energies of the $%
\beta $ and $\gamma $ vibrational states in the rotational bands in $%
^{160}Dy $ are actually not identical, we include a pairing term in (1) for
their calculation, find the parameter $k^{\prime \prime }=14.6keV$ by the
least squares method, and calculate most of the energy spectrum of $^{160}Dy$
positive-parity levels presented in Table 2 in comparison with the
experimental data. Analysis of Table 2 shows that the approach used allows a
satisfactory description of only the lowest excited positive-parity states
in the $^{160}Dy$ nucleus. The discrepancy between the calculated and
experimental level energies considerably increases with increasing
excitation energy. Therefore, we confined ourselves to calculation of
energies of levels with spins $\mathbf{I}_{i}^{\pi }$ with $i\leq 5$, though
states with higher $i$ are experimentally known (for instance, for levels
with $\mathbf{I}^{\pi }$ $=2^{+}$ $i$ may be even larger 16, see Table 1).
In general, the difference between the experimental and theoretical energies
of the states included in Table 2 vary from a few $keV$ to a few hundred
keV. The average deviation of theory from experiment is $<\mid Ee-Ec\mid
>=209.6keV$, which can hardly be regarded as satisfactory. Obviously, this
value will increase if we take into consideration higher-lying excited
states with $i>5$. It should be stressed, however, that the model IBM-1
makes it possible to calculate energies of all head states of rotational
bands which have to be described by individual free parameters in other
models.

\section{DESCRIPTION OF ROTATIONAL BANDS}

We used four widely known, well-working model approaches to describe the
experimental energy spectra of excited states of rotational bands in the $%
^{160}Dy$ nucleus. First of all, it is the geometrical Bohr - Mottelson
model [3], where intraband state energies as a function of the spin $I$ and
the quantum number $K$ are calculated by formula 
\begin{equation}
E_{I}=E_{0}+\sum_{n=1}A_{n}[I(I+1)]^{n}+(-1)^{I+K}\frac{(I+K)!}{(I\_K)!}%
\sum_{m=0}B_{m}[I(I+!)]^{m},
\end{equation}%
where $A_{n}$ and $B_{m}$ are the model parameters.

Another approach is the $Q$ - phonon model [5], where positions of the
rotational-band levels are calculated by expression 
\begin{equation}
E_{I}=E_{0}+\frac{b_{1}}{2}I+\frac{b_{2}}{8}I(I-2)+\frac{b_{3}}{48}%
I(I-2)(I-4),
\end{equation}%
with the parameters $b_{1},$ $b_{2}$ and $b_{3}$.

The third approach is calculation by three-parameter formula following from
the model of the variable moment of inertia with dynamical asymmetry [6]%
\begin{equation}
E_{I}=E_{0}+a_{1}I+a_{2}I^{2}+a_{3}I^{3}+b_{0}(-1)^{i},
\end{equation}
where the parameters $a_{1},$ $a_{2}$ and $a_{3}$ are governed by the moment
of inertia of the ground state of the nucleus and by its "softness" and
asymmetry parameters and where $b_{0}(-1)^{i}$ is the sign-changing term
allowing calculation of bands with any value of the quantum number $K$ $%
(i=(I+1)/2$ at $\Delta I=2$ and odd $I$; $i=(I+2)/2$ at $\Delta I=2$ and
even $I;$ $i=I+1$ at $\ \Delta I=1)$.

Finally, the fourth approach is calculation by formula proposed in [7]%
\begin{equation}
E_{I}=E_{0}+A_{1}I(I+1)A_{2}[I(I+1)]^{2}+A_{1/2}\sqrt{I(I+1)}+B_{0}(-1)^{i},
\end{equation}%
where in addition to the normal terms of the Borh - Mottelson formula, there
appears a term with the parameter $A_{1/2}$ taking into account the Coriolis
interaction and a sign-changing term with the parameter $B_{0}$ similar to
the sign-changing term in (8).

\section{ RESULTS \ AND DISCUSSION}

According to the data collected in [1], there are about 15 known bands of
different nature in $^{160}Dy$, which where established in various types of
nuclear reactions and in the $\beta $ decay. Some of these bands are traced
to rather high energies and spins. For example, in [8] the study of the
reaction in the beam of $^{7}Li$ ions revealed excitation of levels up to
the energy of $7231keV$ with the spin $I^{\pi }=28^{+}$ in the $K^{\pi
}=0^{+}$ ground-state band, up to $6642keV$ \ with $I^{\pi }=25^{+}$ in the $%
K^{\pi }=2^{+}$ $\ S$ band, up to $4875keV$\ \ with $I^{\pi }=20^{+}$ in the 
$S$ band, and up to $6967keV$ with $I^{\pi }=26^{-}$ and $4937keV$ with $%
I^{\pi }=19^{-}$ in the $K^{\pi }=2^{-}$ and $K^{\pi }=1^{-}$ octupole bands
respectively. In our calculations we confined ourselves to the intraband
states with energies not higher than the $^{160m,g}Ho\rightarrow $ $^{160}Dy$
$\beta $ decay energy of $3300keV$. In Tables 3.1-3.16 we present the
energies of excited states in all experimentally established rotational
bands of the $^{160}Dy$ nucleus calculated by (6), (7), (8), and (9) in
comparison with the experimental data. The last two rows of each table show
the average deviations of the calculated energies from the experimental
values $<|Ee-Ec|>$ for each formula and the values of the parameters at
which the best agreement between theory and experiment was achieved. The
second columns of the tables show the values of the inertia parameters $%
\frac{\hbar ^{2}}{2\Theta }$ calculated by the formulae

\bigskip 
\[
\frac{\hbar ^{2}}{2\Theta }=\frac{E_{\gamma }}{4I-2}\text{ \ \ for }%
E(I)\rightarrow E(I-2)\text{ \ transitions, \ \ \ \ \ \ \ \ \ \ \ \ \ \ \ \
(10)}
\]

$\ \ \ \ \ \ \ \ \ \ \ \ \ \ \ \ \ \ \ \ \ \ \ \ \ \ \ \ \ \ \ \ \ \ \ \ \ \
\ \ \ \ \ \ \ $%
\[
\frac{\hbar ^{2}}{2\Theta }=\frac{E_{\gamma }}{2I}\text{ \ \ for \ }%
E(I)\rightarrow E(I-1)\text{ \ transitions.\ \ \ \ \ \ \ \ \ \ \ \ \ \ \ \ \
\ \ \ \ (11)}
\]

Notes $^{a,b,c}$ in Tables 3.1-3.16 are explained under Table 3.16. After
the calculation of the average deviations $<|Ee-Ec|>$ all calculated state
energies were rounded off to the  first decimal digit. It should be
particularly mentioned that when calculating rotational bands for which the
number of the experimentally known levels was not enough, we artificially
extended the band by adding states with approximately expected energies and
obviously large errors. Then we repeated the fitting procedure using the
energies of the missing band levels found in the first fitting. Though
reducing to zero average deviations of theory from experiment for the known
states in some cases, this procedure predicts to an extent positions of
possibly existing but not yet experimentally found intraband levels.

\subsection{ $K^{\protect\pi }=0^{+}$ ground-state band}

Levels of this band are known from the experiment [8] up to $I^{\pi }=28^{+}$%
. We confine our consideration to states with $I^{\pi }\leqslant 16^{+}$
(see Table 3.1), which were studied before the investigation [8] in many
types of nuclear reactions, Coulomb excitation, and $\beta $ decay [1]. The
lowest states with $I^{\pi }=2^{+},4^{+}$ and $6^{+}$ manifest themselves
practically in all above-mentioned processes while the levels with higher
spins $8^{+},10^{+},12^{+},14^{+}$ and $16^{+}$ show up only in some of them
and not in the $\beta $ decay (see [1] for details), where excitations of
these states are unlikely or absolutely impossible because of their high
spins and energies. Recently [2] the state with the energy of $966.8keV$ and 
$I^{\pi }$ $=8^{+}$ was nevertheless found during the investigation of the $%
^{160}Er\rightarrow $ $^{160m,g}Ho\rightarrow $ $^{160}Dy$ decay. This state
is de-excited to the level at $581.1keV$ with $I^{\pi }$ $=6^{+}$ by an
intraband $\gamma $ transition of energy $385.7keV$ which showed itself in
the spectrum of $\gamma \gamma $ coincidences with the $297.5keV$ $\ \gamma $
line de-exciting the $581.1keV$ state. The $966.8keV$ \ level is populated
from higher states by a few low-intensity $\gamma $ transitions whose total
intensity is totally counterbalanced by the intensity of the $385.7keV$ \
transition. As is evident from Table 3.1, our calculations for this band by
all four formulae show approximately the same good agreement with the
experiment. In all cases the average deviation of the experimental energies
from the theoretical ones $<|Ee-Ec|>$ \ does not exceed $7keV$. To get
comparable agreement of the calculations by (8) and (9) with the
calculations by (6) and (7), we had to increase the number of the parameters
from three to four. The inertia parameter $\frac{\hbar ^{2}}{2\Theta }$
tends to decrease smoothly from $14.46$ to $9.30keV$ as the energies and
spins of the band levels increase from $86.8(2^{+})$ to $3091.7(16^{+})keV$.

\subsection{ $K^{\protect\pi }=2^{+}$ $\protect\gamma $ vibrational band}

The first three states of this band (see Table 3.2) with $I^{\pi }$ $%
=2^{+},3^{+}$ and $4^{+}$ are quite well known both from the reactions and
Coulomb excitation and from the $\beta $ decays of the $^{160}Ho$ and $%
^{160}Tb$ nuclei [1]. The $I^{\pi }$ $=5^{+}$ level unambiguously manifests
itself in the $^{160}Ho$ and $^{160}Tb$ $\beta $ decays and in reactions
with $\alpha $ particles. The $I^{\pi }$ $=6^{+}$ state is observed in the $%
^{160}Ho$ $\beta $ decay, in reactions with $\alpha $ particles, and in the
Coulomb excitation. The next six levels with $I^{\pi }$ $%
=7^{+},8^{+},9^{+},10^{+},11^{+}$ and $12^{+}$ were earlier observed only in
reactions with $\alpha $ particles; three of them with $I^{\pi }$ $%
=7^{+},8^{+}$ and $9^{+}$ have been recently confirmed in [2], where the
decay $^{160}Er\rightarrow $ $^{160m,g}Ho\rightarrow $ $^{160}Dy$ was
studied. These states were introduced in the $^{160m,g}Ho\rightarrow $ $%
^{160}Dy$ \ decay scheme on the basis of the $\gamma \gamma $ coincidences
and energy and intensity balances. The last two states with $I^{\pi }=13^{+}$
and $14^{+}$ in Table 2 were first established in the reaction with $^{7}Li$
in [8], where existence of all other members of the band was confirmed and
the more accurate value $2708.0keV$ was found for the energy of the $I^{\pi
} $ $=12^{+}$ level, earlier known [1] to be a level at $2698.0keV$ . As is
evident from Table 3.2, our calculations reproduce the energies of the
levels from this band in the best way in all cases. The average deviation
from experiment $<|Ee-Ec|>$ is no larger than $7keV$ for calculations by
(7), (8) and (9), and $2.67keV$ for the calculations by traditional Bohr -
Mottelson formula (6). In calculations by (6) this agreement was achieved by
including a sign-changing term with the coefficient $B_{0}$, which depends
not only on the spin $I$, as in (8) and (9), but also on the quantum number $%
K$. The parameter $\frac{\hbar ^{2}}{2\Theta }$, as in the case of the
ground-state band, generally tends to decrease slightly with increasing
energy and spin of intraband states. Yet, for the neighboring levels with
even and odd spins a systematic difference in values of $\frac{\hbar ^{2}}{%
2\Theta }$ is observed, which increases with increasing energy. The values
of $\frac{\hbar ^{2}}{2\Theta }$ for odd-spin members of the band are larger
than for even-spin ones.

\subsection{ $K^{\protect\pi }=2^{-}$ octupole vibrational band}

The first head level (Table 3.3) of this band with $I^{\pi }=2^{-}$ and the
third one with $I^{\pi }$ $=4^{-}$ are known from the $\beta $ decay of the $%
^{160}Ho$ and $^{160}Tb$ nuclei and from reactions with $\alpha $ particles
and deuterons [1]. The best studied state from this band is the second level
with $I^{\pi }$ $=3^{-}$, which is easy to observe in both $\beta $ decays,
at the Coulomb excitation, and in almost all reactions mentioned in [1]. The
level with $I^{\pi }$ $=5^{-}$ shows itself in reactions with $\alpha $
particles and deuterons, at the Coulomb excitation, and in the $\beta $
decay of the $^{160}Ho$ nucleus. Higher-lying states with higher spins,
beginning with $I^{\pi }$ $=6^{-}$ and up to $I^{\pi }$ $=14^{-}$, except $%
I^{\pi }$ $=13^{-}$, were observed only in reactions with $\alpha $
particles. It is only recently that excitation of $^{160}Dy$ states with $%
I^{\pi }$ $=6^{-},7^{-},8^{-}$ and $9^{-}$ in the $\beta $ decay has been
observed during the investigation of the $^{160}Er\rightarrow $ $%
^{160m,g}Ho\rightarrow $ $^{160}Dy$ decay [2]. The missing level with $%
I^{\pi }$ $=13^{-}$ showed up in the reaction with $^{7}Li$ [8]. Level
energies calculated for this band are in slightly poorer agreement with the
experimental values than for the bands considered above. In all cases the
average deviation $<|Ee-Ec|>$ is as a large as a few tens of keV (see Table
3.3). The inertia parameter $\frac{\hbar ^{2}}{2\Theta }$ shows considerable
difference in value for states with even and odd spins typical of octupole
bands.

\subsection{ $K^{\protect\pi }=0^{+}$ $S$ band}

This band deserves particular attention because of unusually small spins of
its experimentally observed low-lying levels. The state with $I^{\pi }$ $%
=4^{+}$ (see Table 3.4) was established in the $\beta $ decay of the $%
^{160}Ho$ nucleus and in the reaction with $^{3}He$ nuclei, the level with $%
I^{\pi }$ $=6^{+}$ is known from the $\beta $ decay of the $^{160}Ho$
nucleus and reactions with $\alpha $ particles, the state $I^{\pi }$ $=8^{+}$
is found in the reactions with $\alpha $ particles and $^{3}He$, and the
state with $I^{\pi }$ $=10^{+}$ is found only in reactions with $\alpha $
particles [1]. All these states are assigned to the same band and are
interpreted [1] as members of the band of the states aligned in such a way
that the rotational moment and the moment of two aligned neutrons $i13/2$
are not parallel, as is typical of many known $S$ bands in other nuclei.
Recently the band was extended to $I^{\pi }$ $=20^{+}$ in [8]. We included
in Table 3.4 only two states with $I^{\pi }$ $=12^{+}$ and $14^{+}$ out of
all those observed in [8]. In [1] a $K^{\pi }=0^{+}$ band based on the $%
I^{\pi }$ $=0^{+}$ state with the energy $1443.7keV$ is reported. This band
comprises two more levels with $I^{\pi }$ $=2^{+}$ and $4^{+}$ and energies $%
1518.8$ and $1703.2$ $keV$ respectively. The $I^{\pi }$ $=2^{+}$ level is
known from the $^{160}Ho$ decay and the $(t,p)$ reaction. It is reliably
confirmed in [2] as the $I^{\pi }$ $=2^{+}$ level with the energy $1518.4keV$
while the statement that there exists the $I^{\pi }$ $=4^{+}$ level is based
on 30-year-old data on the $\beta $ decay of $^{160}Ho$ and may be
erroneous. The existence of the $1443.7keV$ head state is not confirmed in
[2] either. The $1357.0keV$ \ $\gamma $ transition associated with this
state on the basis of earlier data is unambiguously placed elsewhere in the $%
^{160m,g}Ho\rightarrow $ $^{160}Dy$ decay scheme on the basis of $\gamma
\gamma $ coincidences in [2]. The $E0$ transition to the ground state, which
was earlier the main evidence for existence of the excited $0^{+}$ level at $%
1443.7keV$, was not observed at all despite our specific search for lines
corresponding to this transition in the spectrum of internal conversion
electrons from the $^{160}Ho$ $\beta $ decay [2]. On the contrary, we found
a state at $1456.7keV$ de-excited by two $\gamma $ transitions. First to the 
$I^{\pi }$ $=$ $2^{+}$ level of the ground band and second to $I^{\pi }$ $=$ 
$2^{+}$ level$\ $of $\gamma $ vibrational band. This state $(1456.7keV)$
been populated by a few $\gamma $ transitions from higher-lying levels with $%
I^{\pi }$ $=1^{-}$, except the $2896.3keV$ level with $I^{\pi }$ $=2^{+}$.
Earlier the state at this energy $(1457keV)$ was observed in $(t,p)$
reactions in [9], where the authors assigned the characteristics $I^{\pi }$ $%
=0^{+}$ to this level and assumed that it and the $2^{+}$ state at $1513keV$
observed by them, which probably corresponded to the $1518.4keV$ \ level
(see Table 3.4), may be assigned to the $S$ band. Though it seems somewhat
strange that there is no noticeable $E0$ transition (our estimation is $%
X=B(E0)/B(E2)<2.2\times 10^{-3})$ from the $1456.7keV$ level to the ground
state, the results of our latest investigations [2] are not in conflict with
the assignment of the spin-parity $I^{\pi }$ $=0^{+}$ to this level. Thus,
if one accepts the interpretation [9], to which we are also inclined, the $S$
band appears to be as shown in Table 3.4, while the data given in [1] on the 
$K^{\pi }=0^{+}$ band built upon the $1443.7keV$ level that proved not to
exist, including the fact that one of its member $(I^{\pi }=4^{+}$ level at $%
1703.2keV)$ is absent, seem to be incorrect. However, it may as well be
hypothesized that the band in question is not an $S$ band but a normal band
built upon the $0^{+}$ state at $1456.7keV$ \ that is similar to other $%
I^{\pi }=0^{+}$ bands established in the $^{160}Dy$ and other nuclei.
Whichever interpretation is true, this band is rotational and we calculated
the energies of its levels by the same rotational formulae (6), (7), (8),
and (9) which we used for other bands considered in this paper. The results
of the calculations are given in Table 3.4. They are in rather good
agreement with the experimental data. In all cases the average deviation $%
<|Ee-Ec|>$ varies between $15$ and $20keV$. However, it should be noted that
equivalent agreement between experiment and calculations was obtained with
only one parameter used in calculations by (6) in contrast to
three-parameter calculations by (7), (8), and (9). The values of the inertia
parameters $\frac{\hbar ^{2}}{2\Theta }$ for the lowest states with $I^{\pi
}=2^{+},4^{+},6^{+}$ $,$ and $\ 8^{+}$ show some irregularity, which is
probably due to different influence of the neighboring bands on the
above-mentioned states. As the energy and spin increase and the influence
becomes weaker or the same for all states, the parameters take on
approximately identical values around $7$ $keV$ (see Table 3.4).

\subsection{ First $K^{\protect\pi }=1^{+}$ band}

In [1] this band is treated as a band upon the two-particle state at $%
1804.7keV$ with $I^{\pi }=1^{+}$ and the $(n5/2[523]-n3/2[521])$
configuration, which also comprises two more levels at $1869.5keV$ with $%
I^{\pi }$ $=$ $2^{+}$ and at $1960.4keV$ with $I^{\pi }=3^{+}$. It is stated
that all three levels were found in the $^{160}Ho$ $\beta $ decay and the $%
1869.5keV$ level also showed itself in $(d,d^{\prime })$ and $(t,p)$
reactions as a state at $1875keV$. In our latest studies of the $\beta $
decay $^{160}Er\rightarrow $ $^{160m,g}Ho\rightarrow $ $^{160}Dy$ [2] we
managed to confirm only the first two states with $I^{\pi }=1^{+}$ and $%
2^{+} $. The previously drawn conclusion that there exists a third level at $%
1960.4keV$ $I^{\pi }=3^{+}$ is likely to be wrong. This state was introduced
on the basis of two $\gamma $ transitions with energies $994.4$ and $%
1873.1keV$ from this level to the $2^{+}$ levels of the $\gamma $ band and
the ground-state band respectively. According to our data [2], the $994.4keV$
$\gamma $ transition is unambiguously placed elsewhere in the $^{160m,g}Ho$ $%
\rightarrow $ $^{160}Dy$ decay scheme and there is no $1873.1keV$ $\ \gamma $
transition at all. Thus, if this band exists, only its first two levels with 
$I^{\pi }$ $=1^{+}$ and $2^{+}$ are reliably established. To extend the
band, we tried to select possible candidates with the necessary quantum
characteristics $I^{\pi }$ $=3^{+},4^{+},5^{+}$ and $6^{+}$ (see Table 3.5)
from the spectrum of the experimentally known (see Table 1) excited states.
The level at $1903.2keV$ \ was earlier known from the $^{160}Ho$ $\beta $
decay and the $(d,d^{\prime })$ reaction, the state at $2049.4keV$ was
observed in the $(t,p)$ reaction as a level of energy $2046keV$, the level
at $2113.7keV$ was first observed in [2] in the $^{160}Er\rightarrow $ $%
^{160m,g}Ho\rightarrow $ $^{160}Dy$ $\beta $ decay, and the state at $%
2187.0keV$ showed itself in the $(d,d^{\prime })$ and $(^{3}He,\alpha )$
reactions as a level of energy $2190$ and $2188keV$ respectively. The common
feature of all these states, including the first two, is existence of $%
\gamma $ transitions de-exciting them to the levels of the ground-state
band. Reproduction of this band by means of the rotational formulae results
in rather good agreement, $<|Ee-Ec|>\symbol{126}$ $4keV$ (see Table 3.5),
but requires a lot of parameters. This is evident from the calculations by
(7), where the average deviation for the formula with fewer parameters is
much worse than $21.33keV$. It is noteworthy that there is a sharp
difference in $\frac{\hbar ^{2}}{2\Theta }$ values between the members of
this band, which is not typical of positive-parity bands.

\subsection{ Second $K^{\protect\pi }=1^{+}$ band}

This band was reported in [11] to be a rotational band built upon the $%
2085.3keV$ level with $I^{\pi }$ $=1^{+}$, a doublet state for the $1694keV$
level with $I^{\pi }$ $=4^{+}$ known as a two-quasiparticle state with
parallel spins of two unpaired nucleons $(n5/2[642]+n3/2[521])$. Apart from
the head level, this band comprised three more levels with $I^{\pi }$ $%
=2^{+},3^{+}$ and $4^{+}$ at $2138.8,$ $2210.2$ and $2286.4keV$
respectively. In [1] the energies $2084.96$ and $2138.14keV$ are given for
two positive-parity states known from the $^{160}Ho$ $\beta $ decay, but
their spins were not uniquely established and their belonging to a band was
not mentioned. To the next two states at $2210.2$ and $2286.4keV$ there
might correspond the levels at $2214$ and $2296keV$ observed in the $(t,p)$
reaction, whose quantum characteristics are not fully known. The latter
levels also shows up in the $(d,t\gamma )$ reaction as a state of energy $%
2294keV$. The states of close energies $2200.8$ and $2309.9keV$ were
observed in our latest studies of the $^{160}Er\rightarrow $ $%
^{160m,g}Ho\rightarrow $ $^{160}Dy$ $\beta $ decay [2], where existence of
the levels at $2084.8$ and $2138.2keV$ was also confirmed. All these four
states are de-excited in a similar way and fit well into the $K^{\pi }=1^{+}$
band under consideration, though their spins are not conclusively
established. This is demonstrated both by the values of the inertia
parameters $\frac{\hbar ^{2}}{2\Theta }$ calculated by us for each member of
the band, which fall within a reasonable range from $10.5$ to $13.5keV$, and
by our calculations by rotational formulae (6), (7), (8) and (9) with the
average deviation of the experimental band level energies from the
theoretical values $<|Ee-Ec|>$ not exceeding $4keV$ even in the worst case
(see Table 3.6). However, there arises a question. Which of the two $K^{\pi
}=1^{+}$ bands considered above (see Tables 3.5 and 3.6) corresponds to the $%
(n5/2[523]-n3/2[521])$ configuration? Based on our theoretical analysis we
believe that preference should be given to the earlier interpretation [11],
where this configuration is assigned to the band built upon the $1^{+}$
state at $2084.8keV$.

\subsection{$K^{\protect\pi }=1^{-}$ octupole vibrational band}

The first four states of this band (see Table 3.7) with $I^{\pi }$ $%
=1^{-},2^{-},3^{-}$ and $4^{-}$ are quite reliably established both from the 
$\beta $ decay and from many reactions [1]. The level with $I^{\pi }$ $%
=5^{-} $ showed itself in the $^{160}Ho$ $\beta $ decay and in the
scattering of deuterons from $^{160}Dy$ nuclei. The states with higher spins 
$I^{\pi }$ $=6^{-},8^{-}$ and $10^{-}$ were previously known only from the
reactions with $\alpha $ particles; excitation of two of them with $I^{\pi }$
$=6^{-}$ and $8^{-}$ was first observed in our recent investigations [2].
Until 2002 year two missing levels with intermediate spins $I^{\pi }$ $%
=7^{-} $ and $9^{-}$ could not be observed in experiments. Only recently [8]
they have first been identified together with the $I^{\pi }$ $=11^{-},12^{-}$
and $13^{-}$ states in the reaction with $^{7}Li$. Our calculations describe
the energies of all members of the band given in Table 3.7 quite
satisfactorily, especially the calculations by (8) and (9), where the
average deviations of theory from experiment are about $5keV$ while in
calculations by (6) and (7) these deviations are worse, $18.1$ and $22.6keV$
respectively. The increase in the number of parameters to five and more in
(6) does not essentially improve the agreement. The values of the parameter $%
\frac{\hbar ^{2}}{2\Theta }$ sharply change as one goes from odd to even
spins of band levels, which is typical of octupole bands.

\subsection{Second $K^{\protect\pi }=1^{-}$ band}

The levels at $2701.1$ and $2720.6keV$ with $I^{\pi }$ $=1^{-}$ and $3^{-}$
(see Table 3.8) were earlier known from the $^{160}Ho$ $\beta $ decay [1]
and were tentatively interpreted in [12] as two-phonon $\beta $ quadrupole -
octupole states. In our latest studies of the $\beta $ decay $%
^{160}Er\rightarrow $ $^{160m,g}Ho\rightarrow $ $^{160}Dy$ [2] we observed
these levels and also states with the excitation energies $2718.9$, $2755.0$
and $2757.1keV$ and respective quantum characteristics $I^{\pi }$ $%
=2^{-},(4^{-})$ and $(4,5)$. All these states, including the levels at $%
2701.1$ and $2720.6keV$, are populated only directly from the $^{160}Ho$
decay and have channels of de-excitation to the levels of the ground-state
band. Therefore, we assume that they may belong to the same band based on
the $2701.1keV$ level with possible characteristics $I^{\pi }K=1^{-}1$.
Calculation by all four formulae describe the energies in the this band
equally well. The average deviation of the experimental energies from the
theoretical values $<|Ee-Ec|>$ varies from $1.22$ to $4.86$ $keV$ (see Table
3.8). Note that the values of the inertia parameters $\frac{\hbar ^{2}}{%
2\Theta }$, though slightly underestimated in comparison with $\frac{\hbar
^{2}}{2\Theta }$ for other bands in $^{160}Dy$, abruptly change as one goes
from even to odd spins, which is typical of negative-parity bands.

\subsection{$K^{\protect\pi }=0^{+}$ band on the $I_{i}^{\protect\pi %
}=0_{2}^{+}$ state}

As follows from [1], the $I^{\pi }=0^{+}$ ground level of this band show
itself in the $^{160}Ho$ $\beta $ decay and the $(p,t)$ reaction, the second
state with $I^{\pi }=2^{+}$ is established in a few types of reactions,
Coulomb excitation, and $^{160}Ho$ $\beta $ decay, and the level with $%
I^{\pi }$ $=4^{+}$ is excited only in the $^{160}Ho$ $\beta $ decay. There
is not doubt about existence of this band, all its members are confirmed in
our latest experiment [2]. The data on the known level energies are well
reproduced by our calculations, except the calculation by (7). These
calculations may help to search for so far unknown but probably existing
states with $I^{\pi }$ $=6^{+}$ and $8^{+}$ in future investigations. The
values of the inertia parameters $\frac{\hbar ^{2}}{2\Theta }$ fall within a
tolerable range (see Table 3.9).

\subsection{ $K^{\protect\pi }=0^{+}$ band on the $I_{i}^{\protect\pi %
}=0_{4}^{+}$ state}

This band is absent in [1] and we are the first to introduce it. During the
investigation of the $^{160}Ho$ $\beta $ decay [10] a state at $1757.2keV$
was observed and assigned the quantum characteristics $I^{\pi }$ $%
=(2^{+},3,4^{+})$ as the most probable in view of the gained data on its
de-excitation. Later this level was confirmed in the investigations of the $%
(t,p)$ reactions [9], where, in addition, a state at $1709keV$ with $I^{\pi
} $ $=0^{+}$ was observed for the first time. Both states distinctly showed
themselves in our recent investigations of the $^{160}Er\rightarrow $ $%
^{160m,g}Ho\rightarrow $ $^{160}Dy$ decay [2], where their energy values
were refined and their quantum characteristics were uniquely determined. The 
$I^{\pi }$ $=0^{+}$ level at $1708.2keV$ is populated by $\gamma $
transitions from four higher states, three with $I^{\pi }$ $=1^{-}$ and one
with $I^{\pi }$ $=2^{+}$. The intensity of these transitions is totally
counterbalanced by the sole $\ E2$ transition of $\ 1621.36keV$ \ to the $%
I^{\pi }$ $=2^{+}$ of the ground band. Quite unexpected, as with the $%
1456.7keV$ $\ 0^{+}$ head level of the $S$ band (see 3.4), is the absence of
the $E0$ transition from the $0^{+}$ level at $1708.2keV$ to the $^{160}Dy$
ground state (our estimation for this transition is $X=B(E0)/B(E2)<3.5\times
10^{-2})$ while for two other $0^{+}$ states at $1280.0$ and $1952.3keV$
known form the $^{160}Ho$ $\beta $ decay $X=0.31$ and $0.14$ respectively.
We regard the state at $1756.9keV$ (the value refined by us, see Table 3.10)
as the second level of the band on the $1708.2keV$ $\ 0^{+}$ level. It is
linked to 16 excited levels with known quantum characteristics via its
populating and de-exciting $\gamma $ transitions for most of which
multipolarities are established [2]. This allowed ambiguity to be removed
and the unique spin-parity $I^{\pi }$ $=2^{+}$ to be established for the $%
1756.9keV$ state. As should be expected, in all cases the calculations
exactly reproduce energies of the first two experimentally known $I^{\pi }$ $%
=0^{+}$ and $2^{+}$ levels with the minimum number of parameters. In a sense
these calculations predict \ the energy range for the search for subsequent
states with higher spins $I^{\pi }=4^{+},6^{+}$ and $8^{+}$ that might exist
in this band (see Table 3.10). The parameter $\frac{\hbar ^{2}}{2\Theta }$
is equal to $8.11keV$, which does not significantly differ from the expected
value.

\subsection{ $K^{\protect\pi }$ $=0^{+}$ band on the $I_{i}^{\protect\pi %
}=0_{5}^{+}$ state}

Two states with $I^{\pi }=0^{+}$ and $2^{+}$ given in Table 3.11 are known
from the $^{160}Ho$ $\beta $ decay [10] and are treated as the members of
the same band [1]. In our latest experiments [2] it was possible to confirm
only the fact that those states existed and to refine their energy values.
As is evident from the table, all calculations describe these states in the
best way and predict possible energies of the states with higher spins which
are not yet found experimentally. As in the case of the previous band, the
parameter $\frac{\hbar ^{2}}{2\Theta }$ has a reasonable value, which is $%
10.07keV$ for this band.

\subsection{$K^{\protect\pi }$\textbf{\ }$=0^{-}$\textbf{\ (octupole?) band}}

This band is treated in [1] as a possible octupole band with $K^{\pi }=0^{-}$
and comprises two experimentally known levels (see Table 3.12). The first
level with $I^{\pi }$ $=1^{-}$ was observed in the $^{160}Ho$ $\beta $ decay
and the $(\gamma ,\gamma \prime )$ reaction, the other level with $I^{\pi }$ 
$=3^{-}$ was observed in the $(d,d\prime )$ reaction and at Coulomb
excitation [1]. Later [2] we observed the $I^{\pi }$ $=3^{-}$ level in the $%
^{160}Ho$ $\beta $ decay as well. The above states are well reproduced by
our calculations, which also yield possible energies of excited states with $%
I^{\pi }$ $=5^{-},7^{-}$ and $9^{-}$ lying higher in energy and belonging to
the band in question (see Table 3.12). The value of the inertia parameter $%
\frac{\hbar ^{2}}{2\Theta }$ is $15.38keV$ and is not in conflict with the
required value.

\subsection{ $K^{\protect\pi }=4^{-}$ $(n5/2[642]+n3/2[521])$ band}

By now four states have been identified in this band (see Table 3.13). The
head states with $I^{\pi }$ $=4^{-}$ is known from the $^{160}Ho$ $\beta $
decay and $(^{3}He,\alpha )$ and $(d,t)$ reactions, the next two levels with 
$I^{\pi }$ $=5^{-}$ and $6^{-}$ showed up in the same processes and in
reactions with $\alpha $ particles, the state with $I^{\pi }$ $=7^{-}$
showed itself only in reaction with $\alpha $ particles [1].

Note that in our latest investigations of the $^{160}Er\rightarrow $ $%
^{160m,g}Ho\rightarrow $ $^{160}Dy$ decay [2] we failed to identify the $%
I^{\pi }$ $=6^{-}$ level at $1954.3keV$. The $665.7keV$ $\gamma $
transition, which according to [1] de-excites this state to the $I^{\pi }$ $%
=5^{+}$ level at $1288.7keV$ and is the main argument in favor of its
existence, should have manifested itself in the spectra of $\gamma \gamma $
coincidences with rather intensive $\gamma $ lines of $1004.9$ and $707.6keV$%
. However, we did not observe anything of the kind. Moreover, our data
dictate two other places for this $\gamma $ transition in the $%
^{160m,g}Ho\rightarrow $ $^{160}Dy$ decay scheme and its multipolarity
composition is predominantly $M1,E2$, which is in conflict with its
placement between the levels of $1954.3$ and $1288.7keV$ with different
parities [2]. As is evident from Table 3.13, the calculations with the same
number of parameters in all cases equally well $(<|Ee-Ec|>\symbol{126}$ $%
2keV)$ describe the energies of four experimentally known states with $%
I^{\pi }$ $=4^{-},5^{-},6^{-}$ and $7^{-}$, and the extension of the band
predicts energies of possibly existing states with $I^{\pi }$ $=8^{-}$ and $%
9^{-}$. The inertia parameters for this band are close in value to those for
other bands in $^{160}Dy$. However, there is no sharp change in the $\frac{%
\hbar ^{2}}{2\Theta }$ values as one goes from even to odd spins as is the
case in other negative-parity bands.

\subsection{ $K^{\protect\pi }=4^{+}$ $(n5/2[523]+n3/2[521])$ band}

Until our latest investigations [2] the band upon the two-particle state
with $I^{\pi }K=4^{+}4$ and the energy $1694.4keV$ was known to comprise two
next levels with $I^{\pi }$ $=5^{+}$ and $6^{+}$ ( see Table 3.14), all of
which, including the head level, were earlier reliably established from the
reactions with $\alpha $ particles and the $^{160}Ho$ $\beta $ decay [1].
The fourth member of this band at $2074.2keV$ with $I^{\pi }$ $=7^{+}$ was
earlier observed in two processes as a level at $2075keV$. In one of them,
the reaction with $\alpha $ particles, it was assigned to the band in
question and in the other, the $(^{3}He,\alpha )$ reaction, it was
interpreted as the head level of the $K^{\pi }$ $=3^{-}$ band. It turned out
that in $^{160}Dy$ there are actually two states at closely spaced energies $%
2074.2$ and $2077.4keV$ but with different quantum characteristics $I^{\pi }$
$=7^{+}$ and $I^{\pi }$ $=3^{-}$, which were unambiguously established in
our recent investigation [2]. It is evident from Table 3.14 that all four
formulae describe the level energies for this band in the best way and also
predict energies of three possibly existing states with higher spins $I^{\pi
}$ $=8^{+},9^{+}$ and $10^{+}$. For the already known levels with $I^{\pi }$ 
$=4^{+},5^{+},6^{+}$ and $7^{+}$ the average deviation of theory from
experiment is no larger than $0.2keV$ in all cases. The parameter $\frac{%
\hbar ^{2}}{2\Theta }$ practically does not change from level to level and
has a reasonable value.

\subsection{ Second $I^{\protect\pi }=4^{+}$ band upon $I^{\protect\pi %
}=4^{+}$}

Only the head state with $I^{\pi }$ $=4^{+}$ was known in this band (see
Table 3.15). It was established in the $^{160}Ho$ $\beta $ decay and the $%
(d,t)$ reaction [1]. Another state at $2194.4keV$ with $I^{\pi }$ $=5^{+}$
was firstly found by us in the recent investigation of the $%
^{160}Er\rightarrow $ $^{160m,g}Ho\rightarrow $ $^{160}Dy$ decay [2].
According to the $\gamma \gamma $ coincidence data, this state, like the $%
2096.9keV$ $I^{\pi }$ $=4^{+}$ head state, is mainly de-excited to the
levels of the $K^{\pi }=2^{+}$ $\gamma $ band, which allows these states to
be regarded as the members of the same band. As might be expected, the
calculations accurately describe the energies of the first two known states
of the band and point to possible positions of the next three levels. The
parameter $\frac{\hbar ^{2}}{2\Theta }$ is $9.76$ $keV$, which is only
slightly different from the values for the $K^{\pi }=4^{+}$ band upon the
state at $1694.4keV$.

\subsection{ $K^{\protect\pi }$ $=3^{-}$ (octupole - vibrational) band}

As was pointed out above, in our work [2] we observed a state at $2077.4keV$
\ with $I^{\pi }=$ $3^{-}$, which seems to correspond to the $2075keV$ state
observed in the $(3He,\alpha )$ reaction and interpreted as the head level
of the $K^{\pi }$ $=3^{-}$ band [1]. We take this interpretation as the
basis and add to this band two more levels at $2143.7keV$ with $I^{\pi
}=4^{-}$ and $2372.4keV$ with $I^{\pi }=6^{-}$ (see Table 3.16), whose
energies and quantum characteristics were first established by us during the
investigation of $^{160}Er\rightarrow $ $^{160m,g}Ho\rightarrow $ $^{160}Dy$ 
$\beta $ decay [2]. The latter level is also known from the $(^{3}He,\alpha
) $ reaction, but only its energy $2372keV$ was found from this reaction.
All three states have a common feature: they are de-excited by $\gamma $
transitions of noticeably intensity to the levels of the $\gamma $
vibrational band [2], which allowed us to regard the states as members of
the same band. As is evident from Table 3.16, our calculations reproduce the
energies of the experimentally known $I^{\pi }=3^{-},4^{-}$ and $6^{-}$
states of this band quite well and predict possible positions of the
intermediate level with $I^{\pi }=5^{-}$ and higher-lying states with $%
I^{\pi }$ $=7^{-},8^{-}$ and $9^{-}$ in the $^{160}Dy$ excitation energy
spectrum. Two values of the parameter $\frac{\hbar ^{2}}{2\Theta }$ for the
even-spin states are closely spaced and do not contradict the expected
values.

\section{CONCLUSIONS}

\bigskip

Most of the experimentally observed energy values of $^{160}Dy$ collective
states levels with positive parity are compared with theoretically
calculated values using interacting bosons model (IBM). The mean differences
between experimental and calculated values are about $210keV$.

As a result of the detail analysis of 16 rotational bans they were
supplemented with 17\ new levels. Also, within this analysis the existence
of $1443.7keV$ $\ 0^{+}$ level was not confirmed, while into$\ ^{160}Dy$
decay scheme was integrated a band with a head level $K^{\pi }=0^{+}$ with
the energy $1708.2keV$ . Large amount of explicit $\beta $ decay states we
could not include into any rotatational band (see Table 1).

The internal states energy values, calculated with phenomenological
equations (6), (7), (8), (9) produce a good agreement with the corresponding
experimental values.

All the models used in our investigation of the levels energies and their
quantum characteristics in very rich and complicate spectrum of $^{160}$Dy
nucleus provide a relatively good agreement with experiment. However, in the
region of high spins and energies the disagreement between calculations and
experimental data increases. So it is somehow straightforward to apply for
our further analysis of this experimental data the recently developed
Interacting Vector Boson Model (IVBM) [13]. This work now is in progress.

The investigation was supported by the RFBR.

\end{document}